\documentclass[twocolumn,10pt]{asme2ej}
\pdfoutput=1

\usepackage{epsfig} 
\usepackage{multirow}
\usepackage{amsmath}

\usepackage{booktabs}
\usepackage{caption} 
\newcommand{\ds}{\displaystyle}
\newcommand{\tb}{\textbf}

\title{An assessment of the two-layer quasi-laminar theory of relaminarization through recent high-Re accelerated TBL experiments}

\author{Rajesh Ranjan
    \affiliation{
	Engineering Mechanics Unit\\
	Jawaharlal Nehru Centre for Advanced Scientific Research\\
	Bangalore, India 560064\\
    Email: rajesh@jncasr.ac.in
    }	
}

\author{Roddam Narasimha\thanks{Address all correspondence to this author.}
    \affiliation{ Engineering Mechanics Unit\\
	Jawaharlal Nehru Centre for Advanced Scientific Research\\
	Bangalore, India 560064\\
    Email: roddam@jncasr.ac.in
    }
}

\begin{document}

\maketitle    

\begin{abstract}
{\it 
The phenomenon of relaminarization is observed in many flow situations, including that of an initially turbulent boundary layer subjected to strong favourable pressure gradients. Available turbulence models have hitherto been unsuccessful in correctly predicting boundary layer parameters for such flows. Narasimha and Sreenivasan \cite{narasimha1973relaminarization} proposed a quasi-laminar theory (QLT) based on a two-layer model to explain the later stages of relaminarization. This theory showed good agreement with the experimental data available, which at the time was at relatively low $Re$. QLT, therefore, could not be validated at high $Re$.   

Some of the more recent experiments report for the first time comprehensive studies of a relaminarizing flow at relatively high Reynolds numbers (of order $5\times 10^3$ in momentum thickness), where all the boundary layer quantities of interest are measured. In the present work, the two-layer model is revisited for these relaminarizing flows with an improved code in which the inner-layer equations for quasi-laminar theory have been solved exactly. It is shown that even for high-$Re$ flows with high acceleration, QLT provides a much superior match with the experimental results than the standard turbulent boundary layer codes. This agreement can be seen as strong support for QLT, which therefore has the potential to be used in RANS simulations along with turbulence models. 
}
\end{abstract}

\begin{nomenclature}
\entry{$\ds U_{e}$}{External velocity}
\entry{$\ds U_{s}$}{Slip velocity in outer layer solution}
\entry{$\ds \rho$}{Density}
\entry{$\ds \nu$}{Kinematic viscosity}
\entry{$\ds P$}{Pressure}
\entry{$\ds \delta$}{Boundary layer thickness}
\entry{$\ds \delta^{\star}$}{Displacement thickness}
\entry{$\ds \theta$}{Momentum thickness}
\entry{$\ds H$}{Shape factor $\ds = \delta^{\star} / \theta$}
\entry{$\ds \tau$}{Shear stress}
\entry{$\ds u_{\tau}$}{Friction velocity}
\entry{$\ds y^{+}$}{Wall distance in viscous units $\ds = \frac{y u_{\tau}}{\nu}$}
\entry{$\ds \kappa$}{von K\'arm\'an constant}
\entry{$\ds K$}{Launder's pressure gradient parameter $\ds = \frac{\nu}{{U_e}^2}\frac{\text{d}U_e}{\text{d}x}$}
\entry{$\ds \Lambda$}{Pressure gradient parameter $\ds = - \frac{\delta}{\tau_{w_0}} \frac{\text{d}P}{\text{d}x}$}
\entry{$\ds \Delta_p$}{Pressure gradient parameter $\ds = \frac{\nu}{\rho u_{\tau}^3}\frac{\text{d}P}{\text{d}x}$}
\entry{$\ds Re$}{Reynolds number}
\entry{$\ds Re_{\theta}$}{Reynolds number based on $\theta$}
\entry{$\ds Re_{\delta}$}{Reynolds number based on $\delta$}
\entry{$\ds x_0$}{Location of beginning of contraction in the experimental set-up}
\entry{$\ds x_1$}{Location of maximum $c_f$}
\entry{$\ds x_{rt}$}{Location of retransition}
\end{nomenclature}
\section{Introduction}
Relaminarization or `reverse transition' occurs when a turbulent or transitional boundary layer reverts to a laminar-like state. This phenomenon was first suspected on a gas turbine blade \cite{wilson1954convective} which exhibited an unexpected drop in the measured heat-transfer co-efficients. Relaminarization has since been investigated not only in the turbine blades  \cite{lagraff2007minnowbrook} but also in many technological flow situations including swept wings \cite{mukund2012multiple}, nozzle contractions \cite{massachusetts1956boundary} and supersonic Prandtl-Meyer expansion corners \cite{sternberg1954transition}; they occur in geo- and bio-mechanical situations as well. It has been observed \cite{narasimha1979relaminarization} that relaminarization of an initially turbulent or transitional boundary layer or duct flow can occur due to one or more of several reasons including: (a) dissipation of turbulent energy due to the action of viscosity or other molecular transport properties, (b) absorption due to an external force or agent such as buoyancy or curvature, and (c) domination of the Reynolds stresses by other imposed forces, such as due to severe acceleration of the free stream in highly favourable pressure gradient (FPG) or a strong normal magnetic field in a magnetohydrodynamic  duct flow. In the first two cases, the decay of turbulent intensity leads to changing the character of the mean flow. However, in the third case, the domination of pressure forces over a nearly frozen or slowly responding Reynolds shear stress plays the major role in relaminarization \cite{narasimha1973relaminarization} (henceforth referred to as NS73). Thus, the Reynolds stresses are not quenched, but their contributions to momentum balance become negligible compared to that of the pressure gradient. It has for this reason been called `soft' relaminarization \cite{narasimha1983relaminarization}, in contrast to the first two cases where there is `hard' relaminarization leading to decay and eventual quenching of turbulence.

\begin{figure}[t]
\centering
\includegraphics[trim=0 0 0 0, clip, width=1.0\linewidth, angle=0]{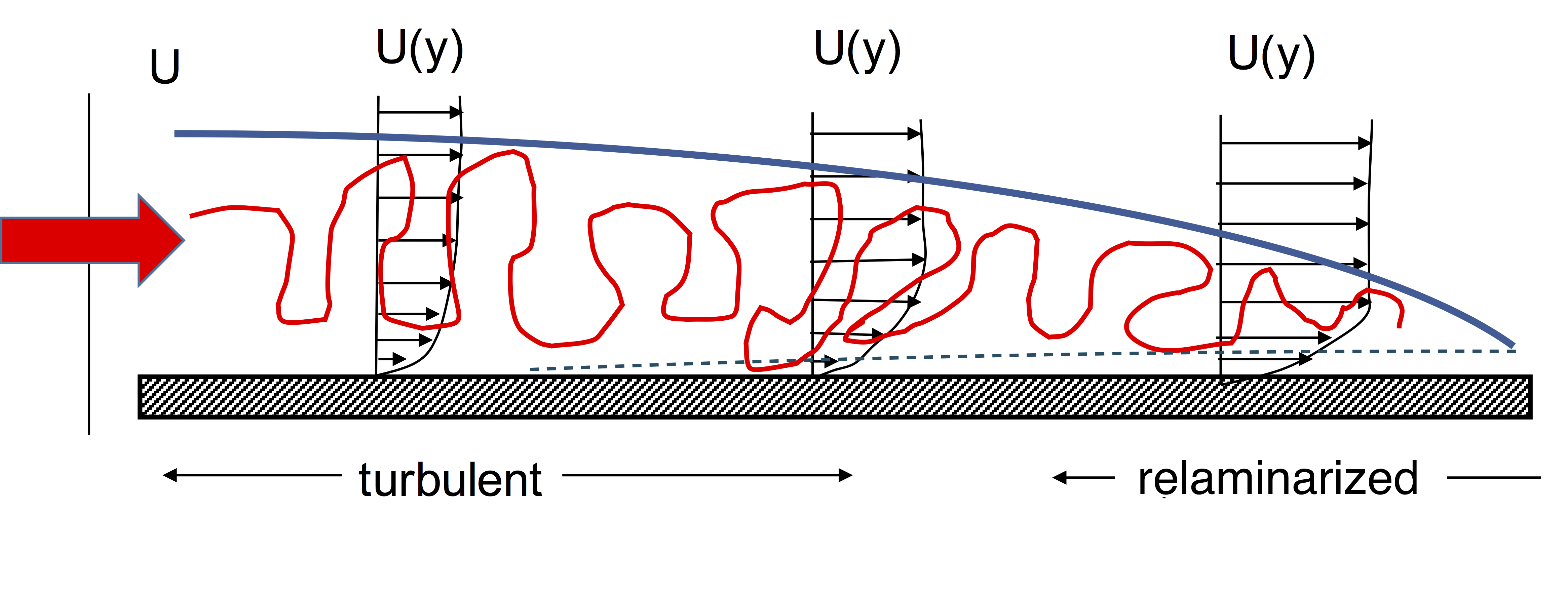}
\caption{`Soft' relaminarization due to strong FPG. Note that the Reynolds stress does not go down in absolute magnitude in the `laminarised' region but become negligible compared to imposed dynamic pressure}
\label{fig:relam}
\end{figure}

In the present work, we will limit our discussions only to relaminarizing boundary layers pertaining to the third category, where the dynamics of reversion is more intriguing.  Figure \ref{fig:relam} is a cartoon of relaminarization due to strong FPG, where one noted characteristic feature of the relaminarizing flow, namely the thinning of the initially turbulent boundary layer, is depicted. It should also be noticed that this relaminarization, unlike direct transition to turbulence (which occurs through the relatively sudden appearance of turbulent spots at a fairly well-defined onset location), is a gradual rather than catastrophic process; e.g. there is no abrupt drop in the boundary layer thickness ($\delta$). It is therefore difficult to define a critical value for some single parameter to identify the onset of relaminarization. Several investigators have nonetheless proposed different indicators of relaminarization and pressure gradient parameters, each with its own critical value, to predict an onset in terms of some flow variable, as listed in Table \ref{tab:parameters} (based on \cite{narasimha1979relaminarization} and \cite{bourassa2009experimental}).  

As depicted in Fig. \ref{fig:relam}, the turbulent intensity does not come down very much in absolute magnitude; rather its normalized value with respect to the free-stream dynamic pressure declines continuously. Because of this the later stages of the relaminarizing boundary layer tends to a state that can be called `quasi-laminar' or `laminarized', as its features are more `laminar-like' than turbulent (in the sense that the mean flow can be predicted without appeal to any turbulence quantity, including the Reynolds shear stress). Warnack \& Fernholz \cite{warnack1998effects}  have compared the mean-velocity profiles with the Falkner-Skan profiles at different streamwise stations in the relaminarizing region, and have noticed that the profiles deviate from the standard `log-law' as we go downstream.  Sufficiently far downstream (in the quasi-laminar region), their match with the Falkner-Skan solutions is reasonably good.  NS73 have also compared the profile of a relaminarized boundary layer with a Blasius profile, and they found the agreement to be excellent. However in both cases turbulent fluctuations remained easily measurable.

Apart from the decrease in $\delta$, the laminarized boundary layer can be easily identified by a rapid increase in the shape-factor ($H$) and a substantial decrease in the skin friction ($c_f$) and heat-transfer co-efficients.  After extensive study of experimental data, NS73 have proposed a quasi-laminar theory (QLT) to predict boundary layer parameters in this  region. This theory has been successfully tested against  experimental data that were available before 1972. These experiments were, however, mostly at relatively low Reynolds numbers ($Re$ of order of a few to several hundreds based on momentum thickness), and often lacked measurements of several crucial boundary layer parameters (e.g. skin-friction) in the same flow. Some of the experimental data available at that time also had considerable scatter or poor momentum balance, and were therefore found to be not sufficiently reliable \cite{sreenivasan1982laminarescent} for validating the theory. 

As experimental as well as DNS data are now available on many relaminarizing flows, at relatively high $Re$ and at high acceleration, we revisit QLT for some of these flows. While these recent investigations confirm the earlier conclusions in general \cite{narasimha1979relaminarization}, none of them attempted quantitative comparisons between the  two-layer theory and the experimental data. It is our objective here to fill this gap. For this purpose, a new improved code has been developed to solve the QLT equations exactly and with higher precision. This study gains significance from the fact that the available turbulence models have not been found reliable for predicting boundary layer parameters in a relaminarizing flow, as will be shown in later sections. 

\begin{table}
\caption{Criterion for the onset of Relaminarization}
\begin{center}
\label{tab:parameters}
\centering 

\begin{tabular}{l|l|l|l}
\toprule
\tb{Ref.} & \tb{Parameter} & \tb{Definition} & \tb{Onset value} \\[6pt]
\midrule

\cite{launder1964laminarization} &$K$ &$\frac{\nu}{U_e^2}\frac{\text{d}U_e}{\text{d}x}$ &$3.0\times10^{-6}$\\[6pt]

\cite{launder1964laminarization} &$H$ &$\frac{\delta^{\star}}{\theta}$ &$ \text{min}(H)$\\[6pt]

\cite{patel1965calibration} &$\Delta_p$ &$\frac{\nu}{\rho u_{\tau}^3}\frac{\text{d}P}{\text{d}x}$ &$-0.025$\\[6pt]

\cite{kline1967structure} &$K$ &$\frac{\nu}{U_e^2}\frac{\text{d}U_e}{\text{d}x}$ &$3.5\times10^{-6}$\\[6pt]

\cite{patel1968reversion} &$\Delta_{\tau}$ &$\frac{\nu}{\rho u_{\tau}^3}\frac{\text{d}\tau}{\text{d}y}$ &$ -0.009$\\[6pt]

\cite{narayanan1969criteria} &$Re_{\theta}$ &$\frac{U_e \theta}{\nu}$ &$300$\\[6pt]

\cite{blackwelder1972large} &$K$ &$\frac{\nu}{U_e^2}\frac{\text{d}U_e}{\text{d}x}$ &$3.6\times10^{-6}$\\[6pt]

\cite{narasimha1973relaminarization} &$\Lambda$& $- \frac{\delta}{\tau_{w_0}} \frac{\text{d}P}{\text{d}x}$&$ 50$\\[6pt]
\bottomrule
\end{tabular}
\end{center}
\end{table}
\section{Recent Studies on Relaminarization}\label{sec:literature}
Sreenivasan \cite{sreenivasan1982laminarescent} has presented an excellent review of experiments on relaminarizing flows available before 1983. He has also identified experiments that can be considered `reliable'  and `trustworthy'. He lamented the fact that despite many experiments whose results were available at that time, there was no single experiment that could be `recommended for turbulence modeling and further computations'. 

Since the publication of that review, there have been at least 4 detailed experimental studies (to the best of our knowledge), which address several concerns raised in \cite{sreenivasan1982laminarescent} through improved setup and instrumentation. We consider that among these studies, the experiment by Bourassa and Thomas\cite{bourassa2009experimental}, apart from having been carried out at the highest initial Reynolds numbers $Re_{\theta0}$ to-date, has implemented most of the suggestions made in \cite{sreenivasan1982laminarescent} for a future experiment, and could be the best single case for developing or validating a model for relaminarization in high FPG. 

\begin{table*}
\caption{Experimental Studies on Relaminarization}
\begin{center}
\label{tab:study_relam}

\def\arraystretch{1.6}
\begin{tabular}{|p{3.2cm}|c|c|c|c|c|c|c|c|c|p{3cm}|}

\toprule
\multirow{2}{*}{\tb{Reference}} & \multirow{2}{*}{\tb{label}} &\multirow{2}{*}{$Re_{\theta0}$}  &  \multicolumn{3}{|c|}{\tb{Contraction Zone}}  & \multicolumn{2}{|c|}{\tb{Max Accel.}} & \multicolumn{2}{|c|}{\tb{Sh. factor}}  & \multirow{2}{2.5cm}{\tb{Setup. Measurement Technique}} \\
\cline{4-10}

& & & $x_0$  & $x_1$ &  $x_{rt}$ & $10^6K$ & $\Lambda$ & $H_{\textrm{min}}$ & $H_{\textrm{max}}$ &  \\

\midrule
 \multirow{2}{3.2cm}{Badri Narayanan \& Ramjee (1969) \cite{narayanan1969criteria}}  & BR2 &310&0.1778&0.2286&0.508  & 8 & 30 & 1.4 & 2.6 & \multirow{2}{3cm}{Wall liner. Pitot. Hot wire. Heat transfer gauge}\\
\cline{2-10}
   & BR3 &410&0.1778 &0.3048& 0.4572& 8.1 & 35 & 1.36 & 2 &  \\
\hline
 Blackwelder \& Kovasznay(1972) \cite{blackwelder1972large}  & BK &  2500 & 9.7 & 9.7& 11.4 &4.8 & 150 & 1.22 & 1.78 &  2D Contraction. Hot-wire, X-probe\\

\hline

\multirow{2}{3.2cm}{Warnack \& Fernholz (1998) \cite{warnack1998effects}}    & WF2 &862&1.218&1.353&1.653& 4 & 63 & 1.34 & 1.68 & \multirow{2}{3.2cm}{Axi. center-body. Preston tube, surface fence, hot-wire, oil-film} \\ [0.5em]
\cline{2-10}
  & WF4 &2564 & 3.238 & 3.483&3.733  & 3.88 & 175 & 1.26 & 1.6 &  \\ [0.5em]
\hline
 Escudier \textit{et. al} (1998) \cite{escudier1998laminarisation}  & ES &1700 & 3.5 & 3.8 &4.5 & 4.4 & - & 1.32 & 2.4 & Contraction. Hot-wire, Hot film\\

\hline
Bourassa \& Thomas (2009) \cite{bourassa2009experimental}  & BT &  4590 &9.14 & 9.21 & 9.67 & 4.5 & - & 1.16 & 1.74 & Contraction. X-wire. OFI \\

\bottomrule
\end{tabular}
\end{center}
\end{table*}

Table ~\ref{tab:study_relam} lists recent laboratory experiments including two entries from the list available in \cite{narasimha1973relaminarization} and \cite{sreenivasan1982laminarescent} which are often quoted in the recent literature. The experiments in which comprehensive data are not available or otherwise considered not `reliable' (according to \cite{sreenivasan1982laminarescent}) are not considered in preparing this list. This table includes a brief summary about each flow including (wherever available) the initial Reynolds number ($Re_{\theta 0}$), beginning of contraction in the experiment ($x_0$), extent of the laminarized zone for the present analysis, defined by the region between maximum $c_f$ near contraction ($x_1 $)  and the minimum $c_f$ at retransition ($x_{rt}$),  maximum acceleration as defined by parameters  $K$ and $\Lambda$, minimum and maximum shape factors as well as a brief description about the set-up and measurement techniques used. Brief comments about these flows are given below.

Badri Narayanan \& Ramjee \cite{narayanan1969criteria} report a series of seven experiments using a tunnel wall-liner as well as wedges to achieve high acceleration. There is usually a large scatter in the data, and skin-friction measurements were made only in two experiments out of the seven. In \cite{narasimha1973relaminarization}, these two cases were considered for the assessment of QLT. The major drawback of these experiments is very low initial Reynolds numbers, which unfortunately do not rule out  low $Re$ effects.     

Blackwelder \& Kovasznay \cite{blackwelder1972large} used a 2D contraction to achieve relaminarization and showed that the Reynolds shear stress in the outer region (away from the wall) remains nearly constant along mean streamlines in the relaminarizing region. This aspect is further discussed in the next section. The intermittency factor however decreased continuously along mean streamlines through strong FPG. Despite being the only detailed study at a relatively high initial Reynolds number ($Re_{\theta0} = 2500$) at the time of Sreenivasan's review, its usefulness is limited by poor 2D momentum balance.

Escudier \textit{et. al} \cite{escudier1998laminarisation} used a contraction shape based on an inviscid analytical solution for a forward-facing step to achieve high acceleration. They used an improved method (compared to BK) to estimate intermittency, and argued that the change in the structure of the turbulent boundary layer can be explained by the fall in intermittency in the vicinity of the wall (from 1 to virtually zero when the flow is fully `laminarized'). In their experiment, it is noticed that as $K$ reaches $K^{\star} = 3\times10^{-6}$ (where superscript denotes the proposed critical value) there is a steep fall in $Re_{\theta}$ from around 1100 to 300 in a distance of 0.1 m - steeper than in any other experiment carried out to-date. Also, Interestingly the region highlighted by \cite{escudier1998laminarisation} in their plots does not correspond to laminarized zone as defined above $(x_1-x_{rt})$ and, in fact,  $c_f$ keeps falling (and $H$ rising) beyond that highlighted region. In the context of these observations, the low-$Re$ effect cannot be ruled out for the sharp and sudden change in boundary layer parameters. Furthermore there is a strong and local departure from two-dimensionality in the region of the steep fall of $Re_{\theta}$.        

Warnack \& Fernholz \cite{warnack1998effects} have performed careful experiments on an axisymmetric center-body in a wind tunnel and focussed on reducing the error in skin-friction measurements due to the use of the same instrumentation in different flow regimes (fully turbulent, laminar-like, zero-pressure gradient etc.). In order to achieve this objective, they have used Preston tubes, surface fences, wall hot-wire probes as well as oil-film interferometry for $c_f$ measurement. Out of the four experiments they performed with this set-up, two (labelled WF2 and WF4 in Table ~\ref{tab:study_relam}) exhibited relaminarization, and confirmed the validity of ideas underlying the analysis of \cite{narasimha1979relaminarization}.

\begin{figure}[t]
\centering
\includegraphics[trim=0 0 0 0, clip, width=1.0\linewidth, angle=0]{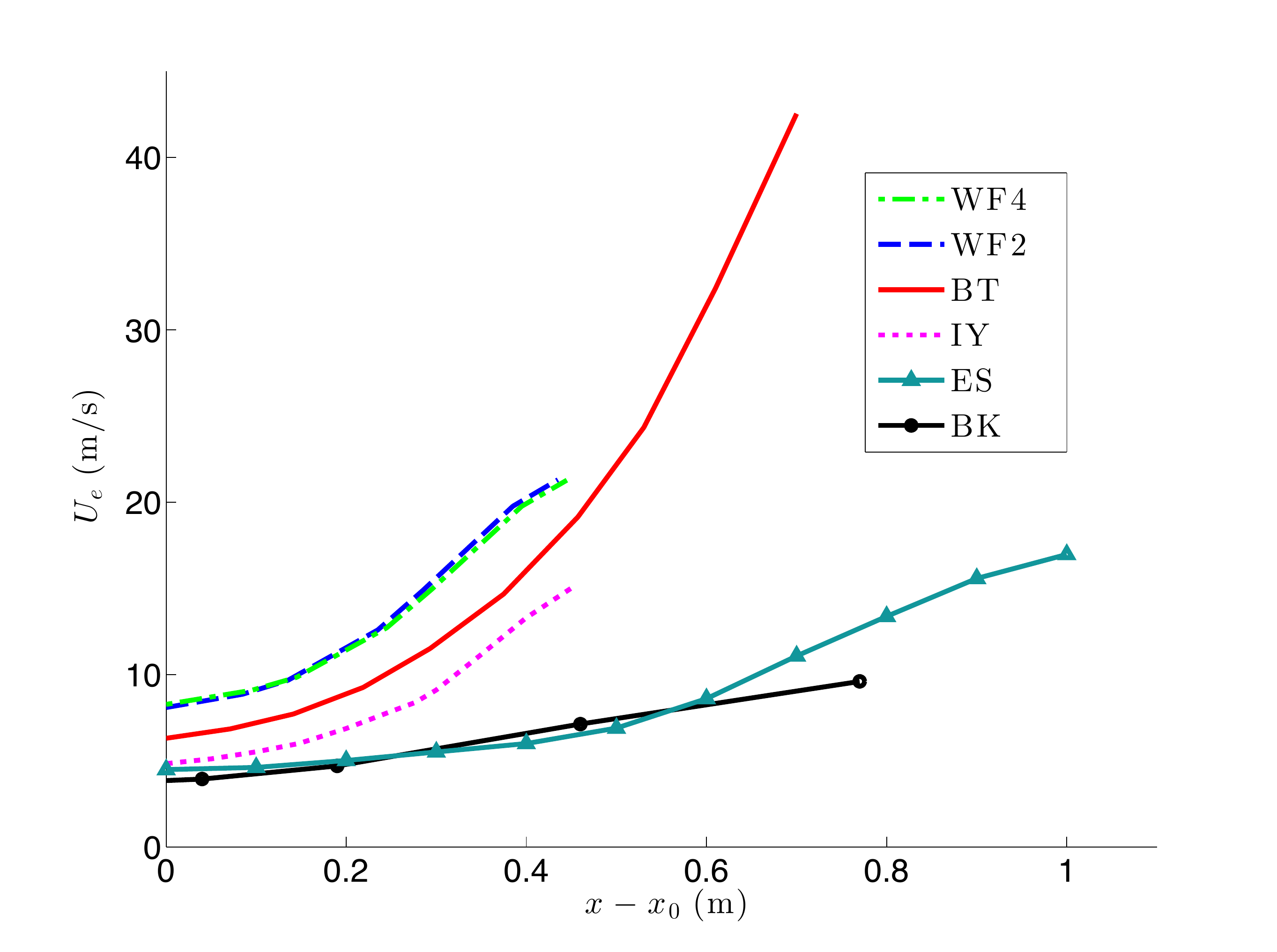}
\caption{Velocity distribution in the relaminarizing region for various flows}
\label{fig:ue_all}
\end{figure}

Bourassa \& Thomas \cite{bourassa2009experimental} present an experimental investigation on a flat-plate turbulent boundary layer in a high contraction which achieves steep favourable pressure gradients over a small distance. The peak value of the pressure-gradient parameter $K$ thus achieved is $4.5 \times 10^{-6}$,  which is 1.5 times higher than the `critical' value $3 \times 10^{-6}$  often considered as necessary for relaminarization.   Compared to other experiments, BT have the longest history of flow development for the `initial' turbulent boundary layer before acceleration is applied, and the highest $Re_{\theta0}$ to-date. In their study, they report most of the flow quantities of interest, measured to high precision. 

Figure \ref{fig:ue_all} shows the external velocity distribution for all the recent experiments with an initial Reynolds number $Re_{\theta0} > 800$ in Table ~\ref{tab:study_relam}. The entry with legend  IY in this figure is taken from \cite{ichimiya1998properties} ($Re_{\theta0} = 799$) . The work of \cite{ichimiya1998properties} is very largely concerned with the structural changes of the relaminarizing boundary layer but no data on mean flow properties are available.  The velocity distribution for WF2 and WF4 are similar as the same centre-body was used to generate the pressure distribution.  Clearly BT stands out from the rest, with the strongest acceleration  ($U_e$ increases by a factor of about 5.14 in a contraction length of 0.61 m). The high strain rate present in this region changes the structure of turbulence enormously and the local skin friction $c_f$ falls by a factor of around 2.5 within the contraction.

In the light of this discussion it can be seen that among available experiments, BT-flow presents the most severe high-$Re$ case for assessing as well as developing relaminarizing flow models. A rigorous assessment of QLT for this flow, as presented in the following sections, is a first step towards development of such models.

Here it should be pointed out that all the experiments mentioned in  Table ~\ref{tab:study_relam}, except BT, report breakdown of the standard log-law  in the relaminarized region. BT, however, argue that the logarithmic law persists even in the relaminarized region, but with substantially different values of slope and intercept. It may require a few more experiments at high $Re$ to settle this issue.  
 
There have also been  more recent DNS studies (\cite{piomelli2013numerical}, \cite{patwardhan2014effect}) which confirm most of the general features of the relaminarizing flow as described here. They are not considered here for the present study as the initial Reynolds numbers ( $Re_{\theta0} = 458$ for \cite{piomelli2013numerical}, and $Re_{\theta0} = 1130$ and 1900 for \cite{patwardhan2014effect}) for these simulations are still appreciably lower than that of \cite{bourassa2009experimental}.

\section{Quasi-laminar theory}
NS73 have proposed a quasi-laminar theory based on their two-layer model, comprising a viscous inner layer and an outer stress-free (hence inviscid) but rotational layer, to explain the mechanics of relaminarization. As the present analysis follows \cite{narasimha1979relaminarization}, only a summary of the method is given below.

In the region with high favourable pressure gradients, the turbulent structures in the outer layer are distorted due to rapid flow acceleration. This leads to the Reynolds shear stress being `frozen' along streamlines and hence out of step with the steep rise in the dynamic pressure. Consequently the  pressure gradient dominates the slowly responding Reynolds shear stress in the outer layer, i.e. $\ds \textrm{d}p/\textrm{d}x \gg \partial \tau/\partial y$, and hence the boundary layer begins drifting away from a fully turbulent state to a turbulence-independent.  Thus the outer layer can be treated as stress-free for calculations, with a slip velocity $U_s$ at the surface.

In order to satisfy the no-slip boundary condition, an inner viscous sub-boundary layer develops subsequently. (In other words, the viscous sublayer already present in the turbulent boundary layer is transformed to a laminar sub-boundary layer.)  This layer can be thought of as  originating from the decaying upstream turbulence in the viscous sublayer of the turbulent boundary layer and is maintained in a stable state by the highly favourable pressure gradient.

\subsection{Validity of QLT for recent experiments}
We now discuss two aspects of QLT: frozen Reynolds stress and  conservation of mean vorticity, in order to ascertain their validity for high $Re$  flows. These aspects, described in \cite{narasimha1973relaminarization} and \cite{narasimha1979relaminarization}, offer further insights into the mechanics of relaminarization.  

\subsection*{Reynolds Stress}
\begin{figure}[t]
\centering
\includegraphics[trim=0 0 0 0, clip, width=1.0\linewidth, angle=0]{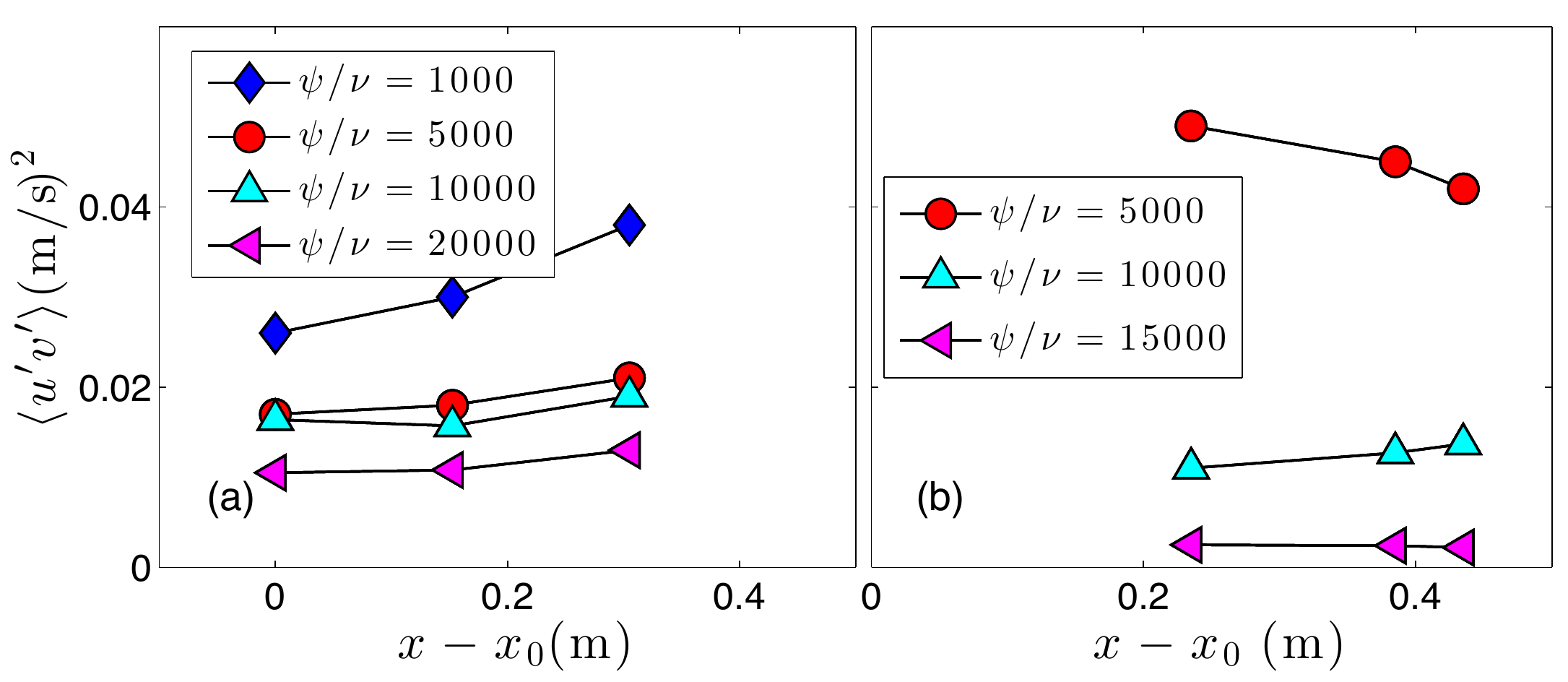}
\caption{Reynolds shear stress along streamlines (a)BT (b)WF2}
\label{fig:uv_all}
\end{figure}

Returning to the Reynolds stresses, the shear stress in the BK-flow was found experimentally to be varying little along streamlines compared to the changes in the dynamic pressure $(1/2) \rho U_e^2$ in the outer layer. A similar observation was made in  \cite{narasimha1979relaminarization} regarding the BR experiments \cite{narayanan1969criteria}. Among the more recent experiments, Reynolds stresses  from the available data for BT and  WF2 (X-wire data for  WF4 are not available), calculated along the mean streamlines, are shown in Figure \ref{fig:uv_all}. The streamfunction was calculated using mean velocity profiles given for these experiments in wall variables $u^+, y^+$ from $\psi/\nu = \int u^+ \textrm{d}y^+$. It can be noted from Fig. \ref{fig:uv_all}  that for BT-flow, as we go away from the wall in the boundary layer (increasing $\psi/\nu$), there is less than 20\% variation in the Reynolds shear stress, whereas the dynamic pressure increases nearly 20 times (see Fig. \ref{fig:ue_all}) over the same distance. 

\subsection*{Conservation of mean-flow vorticity}
One direct consequence of the outer layer being stress-free is conservation of mean angular momentum (hence mean vorticity $\omega \sim (U_e -U_s)/\delta$). This explains the thinning of the boundary layer in the relaminarizing region as the difference between $U_e$ and  $U_s$ decreases as we go downstream (shown in Fig. \ref{fig:BT_ue_us} later). This inviscid nature of the outer layer also restricts entrainment across the layer. NS73 have shown that this was supported by the experimental evidence at the time.

\begin{figure}[t]
\centering
\includegraphics[trim=0 0 0 0, clip, width=0.9\linewidth, angle=0]{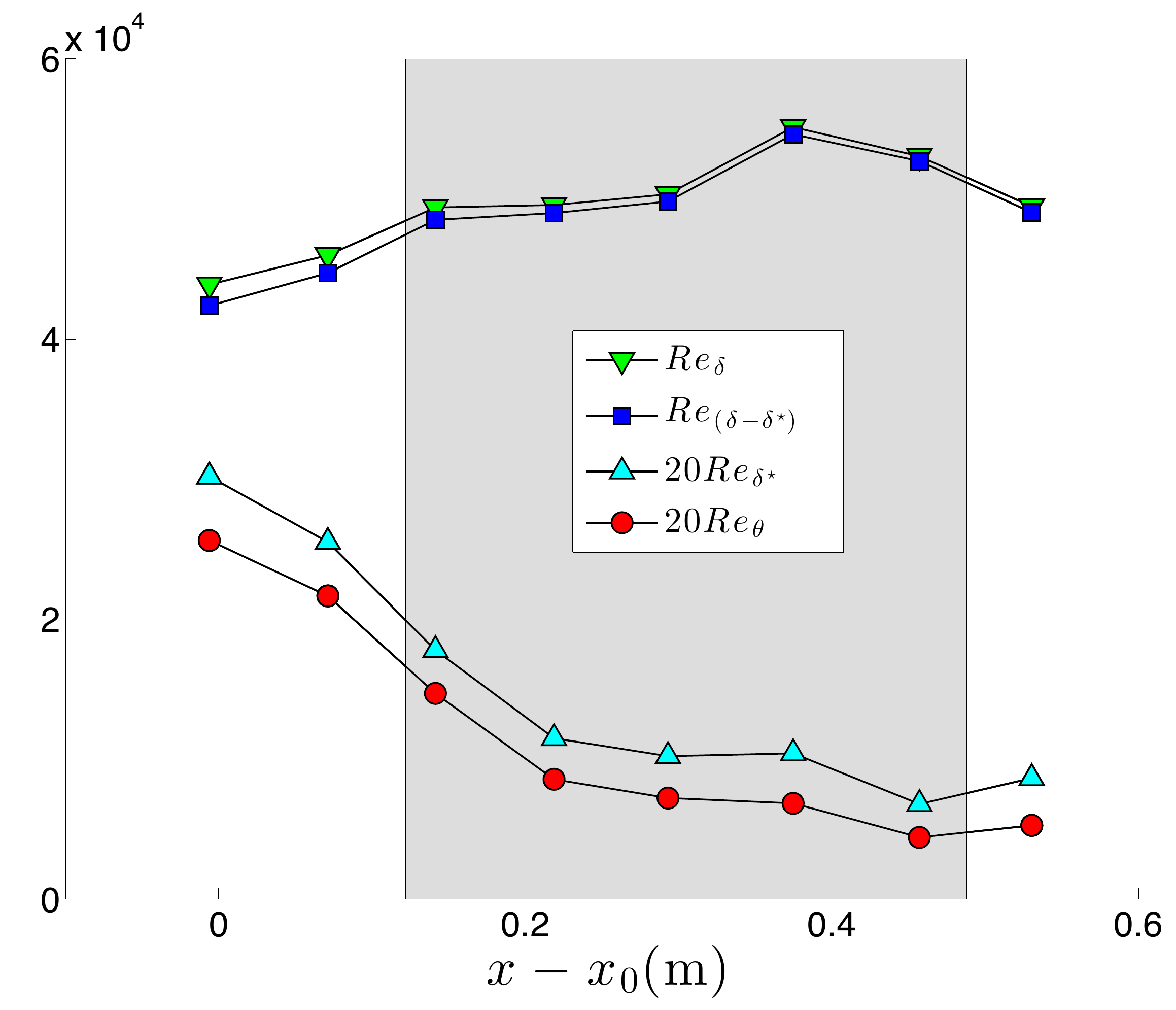}
\caption{Boundary layer Reynolds numbers in the BT-flow. The shaded portion shows the laminarized region ($x_1$ to $x_{rt}$), where QLT is expected to be valid (this convention is followed in the later plots also)}
\label{fig:BTredelta}
\end{figure}

In Fig. \ref{fig:BTredelta}, the Reynolds numbers based on $\delta, \delta^{\star}$ and $\theta$ are plotted for BT-flow. Also included is the Reynolds number based on $(\delta-\delta^{\star})$, which is a measure of the boundary layer mass-flux. It is obvious from the plots that in the laminarized region there is very little variation in $Re_{\delta}$ and  $Re_{(\delta -\delta^{\star})}$, in contrast to $Re_{\delta^{\star}}$ and $Re_{\theta}$. This strengthens the argument of nearly constant mass flux in the boundary layer, and hence very little entrainment. 

\begin{figure}[t]
\centering
\includegraphics[trim=0 0 0 0, clip, width=1.0\linewidth, angle=0]{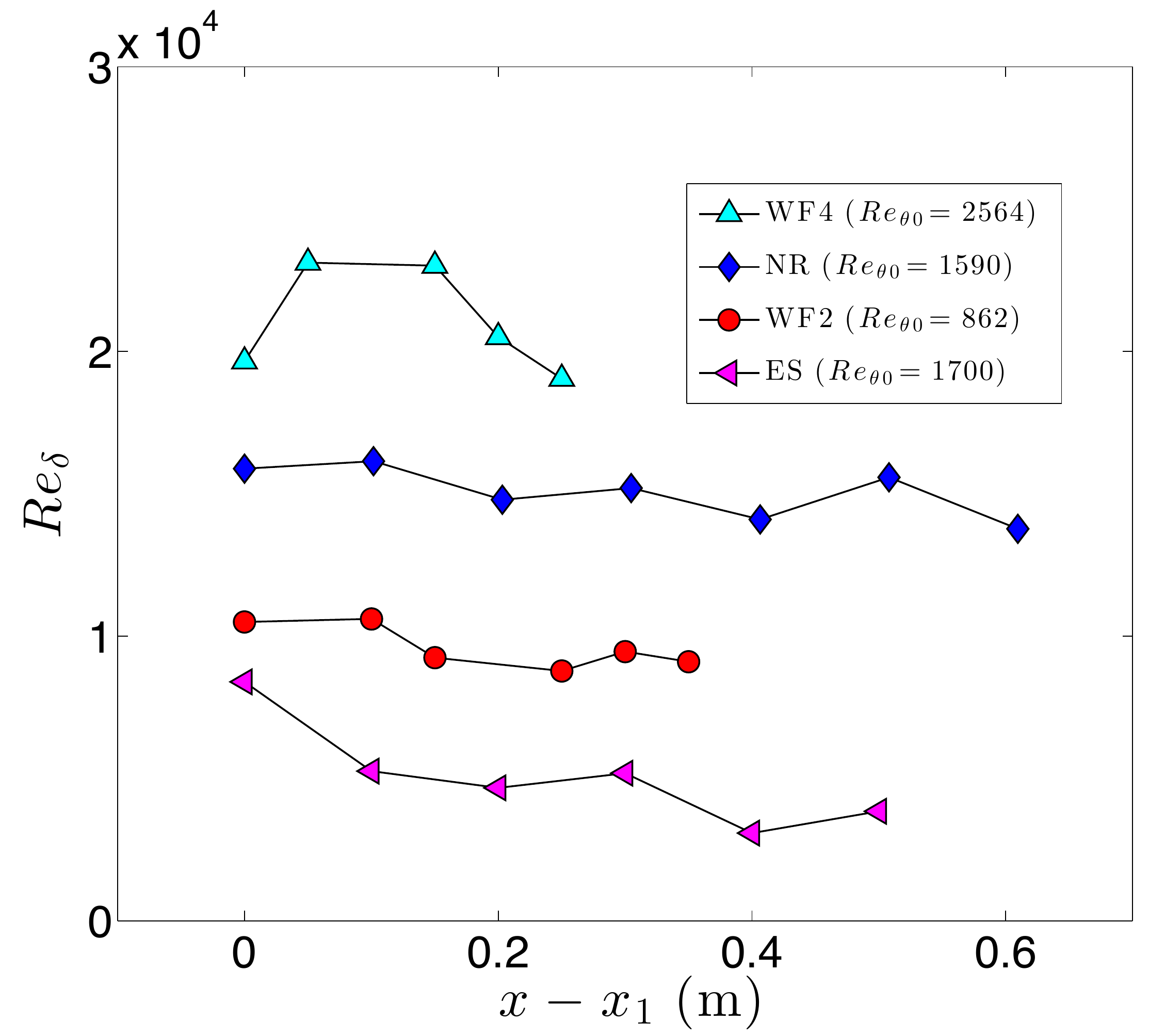}
\caption{Boundary layer Reynolds numbers in the laminarized region in other studies}
\label{fig:redelta}
\end{figure}

This is confirmed in Fig. \ref{fig:redelta}, showing data on $Re_{\delta}$  for other relaminarizing flow experiments.  This plot also includes an entry from an experiment by Narahari Rao (labelled NR) in \cite{krsnotes1974} . As $\delta^{\star}$ is very small compared to $\delta$, $Re_{\delta}$ closely indicates the mass-flux across the layer. Barring a little scatter (for example in WF4), the value of $Re_{\delta}$ remains approximately constant for all streamwise locations in the laminarized region, indicating again little entrainment.

\subsection{Solution of QLT}
\begin{figure}[t]
\centering
\includegraphics[trim=0 0 0 0, clip, width=1.0\linewidth, angle=0]{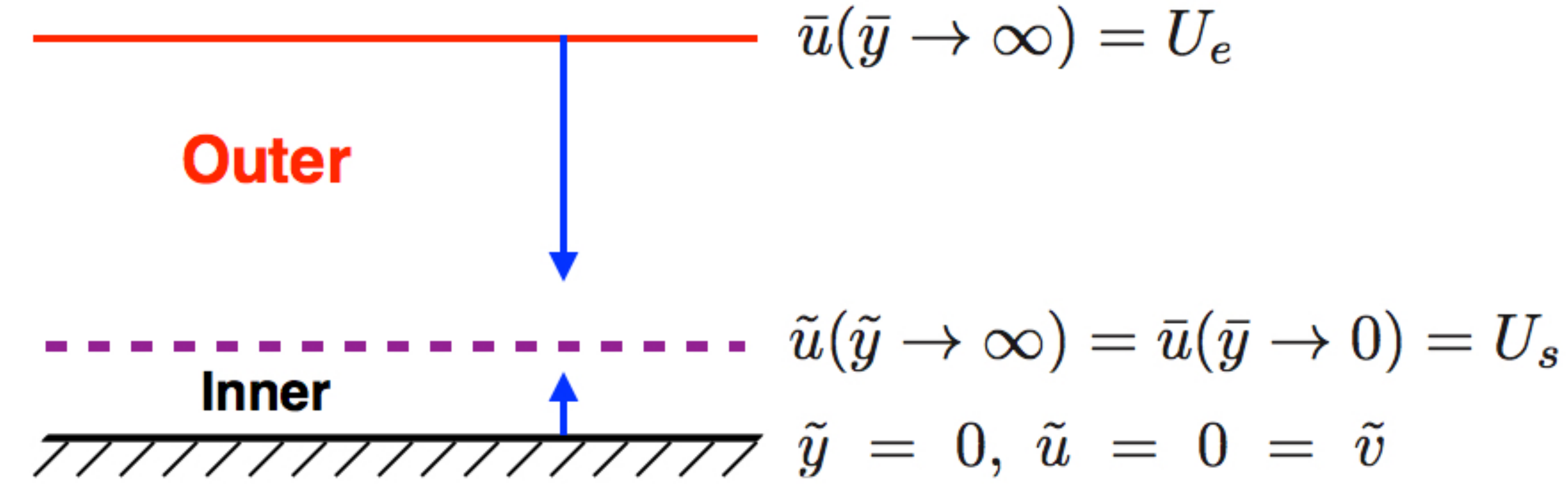}
\caption{Two-layer model. Matched asymptotic solution}
\label{fig:qlt}
\end{figure}
Based on the arguments presented above, a two-layer model was proposed by NS73 to solve for the flow in the quasi-laminar region.  Figure \ref{fig:qlt} depicts these two layers that constitute the quasi-laminar boundary layer. In this model the outer layer is governed by the equation  

\begin{equation}\label{eq:outer}
\ds \bar{u}\frac{\partial{\bar{u}}}{\partial \bar{x}} + \bar{v}\frac{\partial{\bar{u}}}{\partial \bar{y}} = U_e \frac{\textrm{d}U_e}{\textrm{d}\bar{x}}
\end{equation}
and the inner by
\begin{equation}\label{eq:inner}
\ds \tilde{u}\frac{\partial{\tilde{u}}}{\partial \tilde{x}} + \tilde{v}\frac{\partial{\tilde{u}}}{\partial \tilde{y}} = U_s \frac{\textrm{d}U_s}{\textrm{d}\tilde{x}} + \nu \frac{\partial^2{\tilde{u}}}{\partial {\tilde{y}}^2}
\end{equation}
Here the overbar and tilde represent outer and inner variables respectively.

The boundary conditions for each layer are also shown in Fig. \ref{fig:qlt}, based on the matched asymptotic expansions in the two layers (NS73). Here the slip (or surface) velocity $U_s$, which the inner layer sees as the `free-stream' velocity at its edge, can be calculated after the  origin of the inner laminar layer (called `virtual origin') is fixed.

\subsection{Choice of virtual origin}
\begin{figure}[t]
\centering
\includegraphics[trim=0 0 0 0, clip, width=1.0\linewidth, angle=0]{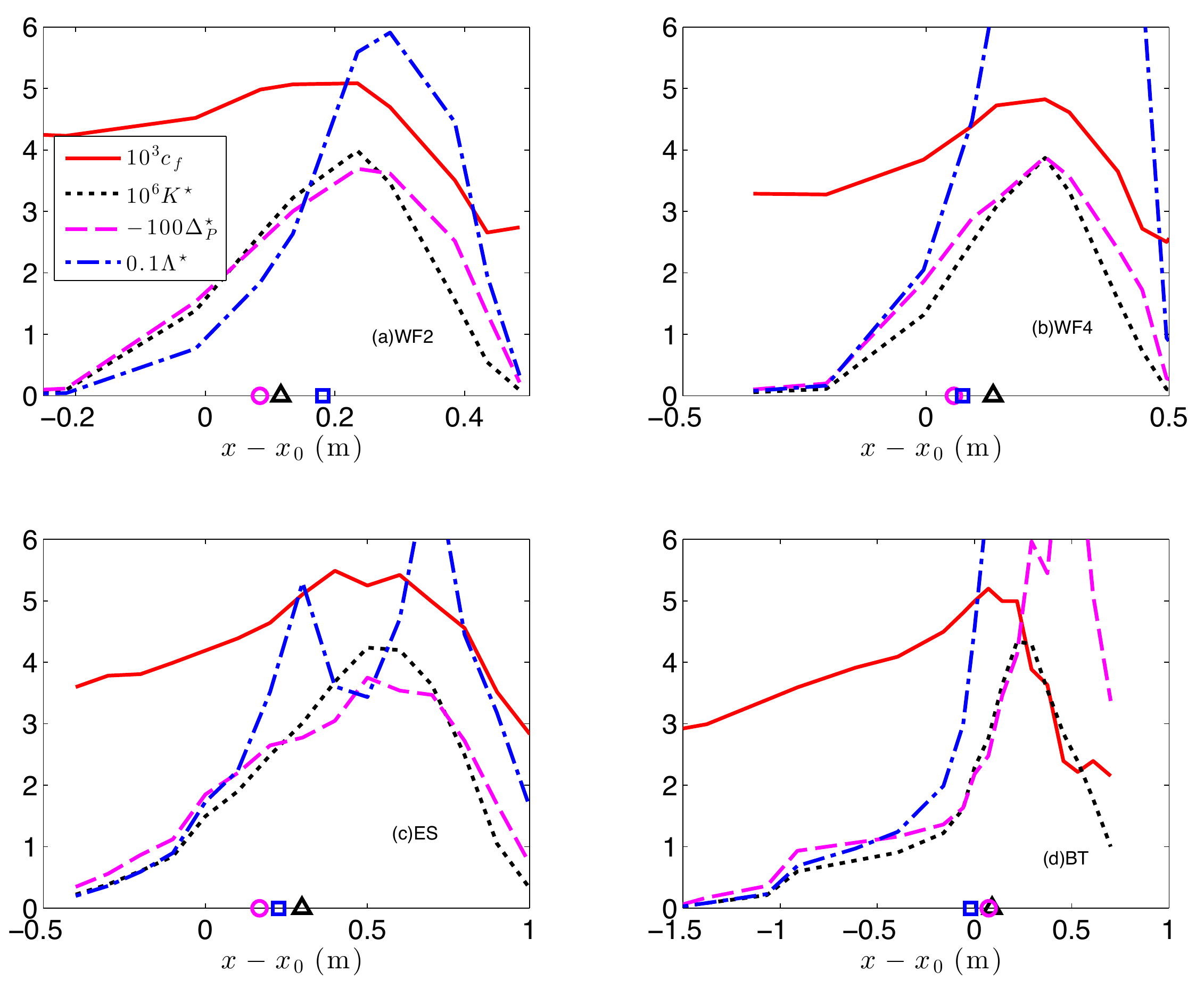}
\caption{Different pressure gradient parameters for the prediction of onset of relaminarization. Triangle - $K^\star$, Circle- $\Delta_p^{\star}$, Square-$\Lambda^{\star}$ }
\label{fig:pgp}
\end{figure}

The precise definition of virtual origin is a grey area, as there is no single parameter to locate a precise `onset of relaminarization'. In Fig. \ref{fig:pgp}, different pressure gradient parameters are plotted over the $c_f$ plot for four flows. The proposed critical values for the parameters $K^{\star}, \Delta_p^{\star}, \Lambda^{\star}$ listed in Table \ref{tab:parameters} (with $K^{\star}$ as suggested by \cite{launder1964laminarization}) are also plotted on the abcissa. There is some scatter among these criteria, but as observed by NS73, a precise definition of the virtual origin is not compulsory for quasi-laminar calculations, as the effect on the solution is not significant unless the origin is appreciably altered. 

This has been largely found to be true for the current calculations where the beginning of the contraction ($x_0$) has been taken as the virtual origin (here $\Delta_p^{\star} \simeq -0.02$).  In the case of BT-Flow, however, quasi-laminar calculations with $x_0$ as virtual origin lead to  slightly over-predicted skin-friction values. However, when the origin is shifted further upstream ($x-x_0 = -1.57$m), the predictions are improved (see Figure \ref{fig:master} in section \ref{sec:results}) . Interestingly there are no appreciable effects of this shift of origin on the other parameters such as  shape factor or boundary layer thickness.   

This shift of the effective origin of the inner laminar layer could be due to the upstream effect of the flow in the BT-experiment, with its long history of flow before the sudden and large contraction compared to other relaminarizing experiments. The upstream effects of this contraction can be seen in Figure \ref{fig:master}, where velocity as well as $K$ are shown to be rising even 1.57 m before the contraction. It has been found that one satisfactory way of determining an effective virtual origin for the laminar sub-boundary layer would be to see if $U_s(x)$ in the laminarised region belongs to the Falkner-Skan family. This can be done by putting $U_s \sim {(x-x_0)}^m$, where both $x_0$ and $m$ can be determined by a least-squares procedure. We have actually done this for the BT-flow and the fit is statistically good to $r^2 = 0.99$, with $x_0 = -1.5$m and $m = 7.4328$. 
   
\begin{figure}[t]
\centering
\includegraphics[trim=0 0 0 0, clip, width=1.0\linewidth, angle=0]{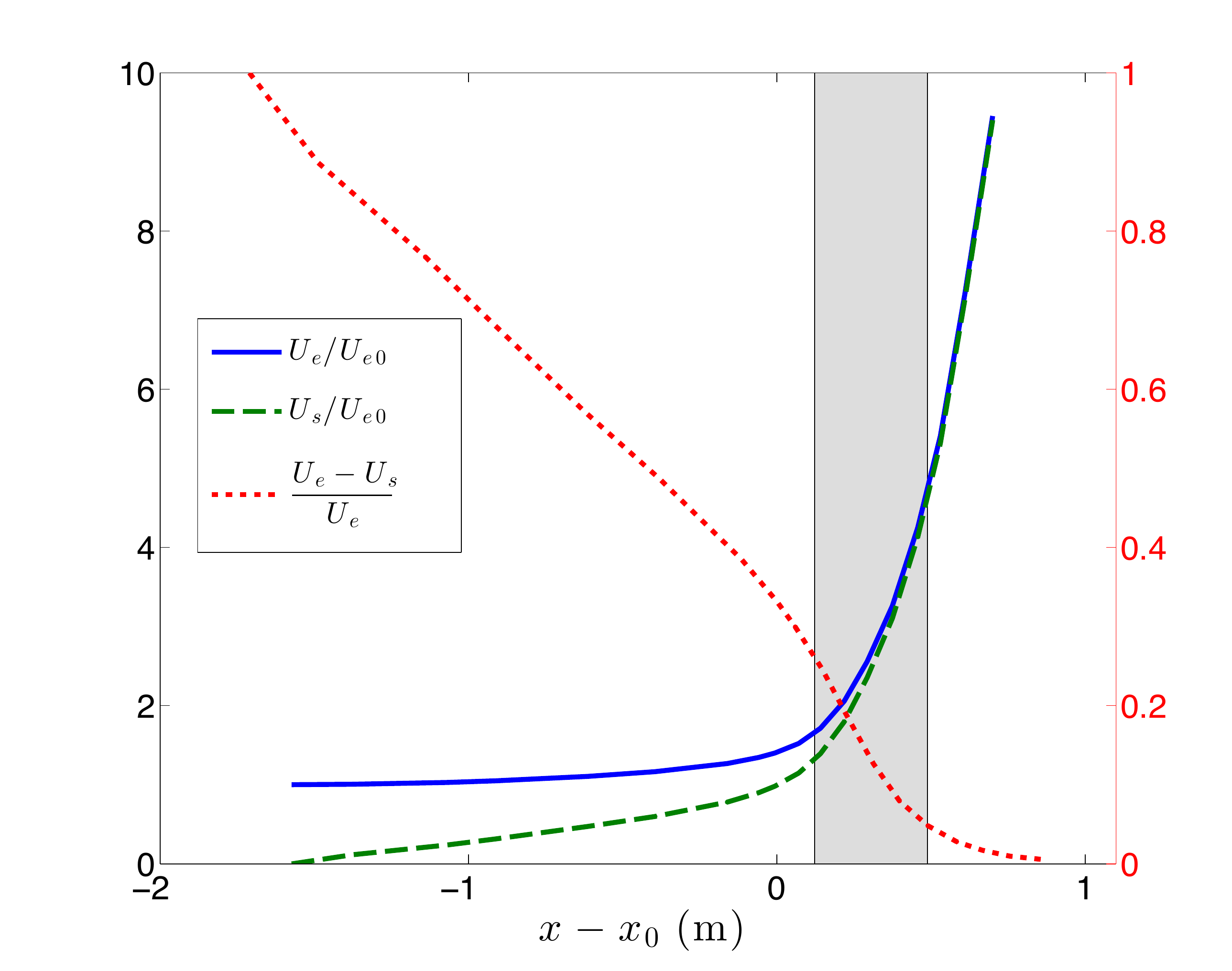}
\caption{External velocities for outer ($U_e$) and inner ($U_s$) layer calculations. Downstream of the relaminarizing region, ($U_e-U_s$) decreases very fast.}
\label{fig:BT_ue_us}
\end{figure}

After the virtual origin is fixed, NS73 computed the slip velocity for the outer layer (or the `external' velocity for the inner layer) using  $U_s = \sqrt{(U_e^2 - U_{e0}^2)+U_{s0}^2}$, where $U_{e0}$ and $U_{s0}$ are respectively external and slip velocities at the virtual origin.  $U_{s0}$ can be taken as zero or a very small value. Figure \ref{fig:BT_ue_us} shows $U_e$ and $U_s$ for the BT-flow.

Once $U_s$ is known, solutions for both the layers can be computed separately using their respective governing equations, and the uniformly valid solution can be obtained by taking the union of the two. 
\section{Numerical Method}
The predictions for the relaminarizing flows have been obtained using both turbulent boundary layer (TBL) models and quasi-laminar theory (QLT). Two separate codes are written to solve their respective 2D equations using an implicit finite difference method. These codes are written in MATLAB and largely follow the schemes described in \cite{schetz1993boundary}.

\subsection{Turbulent boundary layer}

In this code, the full turbulent 2D incompressible boundary layer equations, as given below, are solved.  
\begin{equation}\label{eq:tbl}
\ds u\frac{\partial u}{\partial x} + v\frac{\partial u}{\partial y} = U_e \frac{\textrm{d}U_e}{\textrm{d} x} + \frac{1}{\rho} \frac{\partial}{\partial y}\left( \mu \frac{\partial u}{\partial y}\right) - \rho \overline{u'v'}
\end{equation}
The flow is assumed to be fully turbulent from the first reported location in the experiments (which is the case for the experiments mentioned). The initial skin friction is obtained using the Blasius turbulent skin-friction law and the initial boundary layer velocity profile is formed using the Coles velocity profile combining the law of the wall and the law of the wake. A total of 5000 points were used across the boundary layer without any grid stretching. No-slip  boundary condition was used at the wall without any wall-function model.

\subsubsection{Turbulence models}
The Reynolds shear stress term in Eqn. \ref{eq:tbl} is obtained with a mixing length model (MLM) as well as an algebraic eddy viscosity model (EVM). For MLM, a length scale ($l_m$) based on Prandtl's mixing length idea is introduced to calculate the stress term 
\[ -\rho \overline{u'v'} = \rho l_m^2 \left( \frac{\partial u} {\partial y} \right)^2 \]
whereas in the case of EVM, the eddy viscosity ($\mu_T$)  is introduced to simplify the equations to the form of the laminar flow equations, assuming that the shear-stress is given by: 
\[ -\rho \overline{u'v'} = \mu_T \left( \frac{\partial u} {\partial y} \right) \]
There are numerous models available in the literature to compute $l_m$ and $\mu_T$.  Here we follow the composite models as given in \cite{schetz1993boundary}. The turbulent boundary layer in terms of $l_m$ or $\mu_T$ can be divided in two layers - wall and outer. The model parameters used in our calculations for these two layers are listed in Table \ref{tbl:TBL}.   
\begin{table}
\caption{Turbulence models used in BL calculations}
\begin{center}
\label{tbl:TBL}
\def\arraystretch{1.6}
\begin{tabular}{ccc}
\toprule
 & Outer & Wall \\
\midrule
MLM ($~l_m$) &  $\ds 0.09\delta$ & $\ds \kappa y \left[ 1 - exp(-\frac{y^+}{A^+}) \right] $ \\ [1em]
EVM$~(\mu_T)$  &  $\ds 0.018\rho u_e \delta^{\star}$   & $\ds \kappa \rho \nu \left[  y^+ - y_a^+ \textrm{tanh} \left(\frac{y^+}{y_a^+}\right) \right]$  \\
\bottomrule
\end{tabular}
\end{center}
\end{table}
\begin{figure}[t]
\centering
\includegraphics[trim=0 0 0 0, clip, width=0.9\linewidth, angle=0]{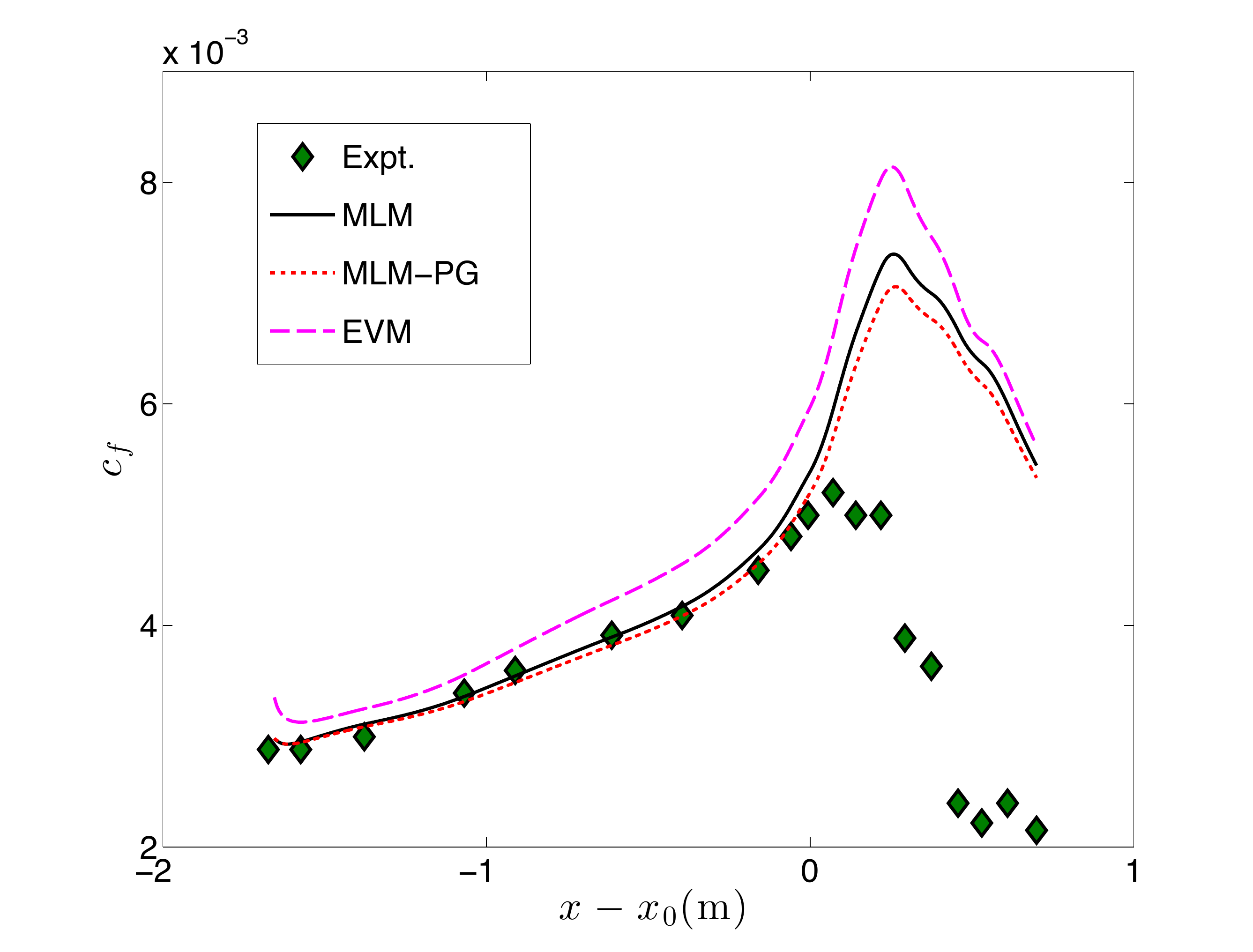}
\caption{Prediction using TBL code for BT-flow}
\label{fig:BT_turbulence}
\end{figure}
Here $\kappa$ is the von K\'arm\'an constant, taken as 0.41 for the present calculations. The wall layer MLM is the well-known Prandtl-Van Driest mixing length model, with damping length constant $A^+ = 26$. Cebeci and Smith\cite{cebeci1968finite}  introduced modifications in this model to account for the pressure gradient in the flow, taking the constant as $A^+ = 26/N$, where $N = (1+11.8\Delta_p)^{1/2}$. 

In EVM, the outer and wall layer models are due to \cite{clauser1956turbulent} and \cite{reichardt1951complete} respectively. Here $y_a^+$ is another dimensionless length scale in the order of laminar boundary layer thickness and is taken as 9.7 (after  \cite{clauser1956turbulent}). 

Figure \ref{fig:BT_turbulence} shows the predictions of skin-friction for BT flow with the different turbulence models described above. In the fully turbulent region where acceleration is not very high ($x < x_0$), both MLM and MLM-PG (MLM with pressure-gradient modifications suggested in  \cite{cebeci1968finite}) show good match with the experimental data but all of them fail conspicuously in the relaminarizing region. MLM-PG  does only slightly better than MLM in this region.

EVM consistently overpredicts the $c_f$ value, even in the region $x<x_0$. This may be because the Clauser model \cite{clauser1956turbulent} is developed mainly for equilibrium pressure gradient turbulent flows, which is not the case for the BT-flow. However, as \cite{schetz1993boundary} concludes, ``there is no other known eddy viscosity model that is more generally applicable".

We have also solved the turbulent boundary layer using several momentum intergral methods including the methods proposed by Spence \cite{spence2012development} (used by \cite{narasimha1973relaminarization}), Head \cite{head1960entrainment} and Moses \cite{moses1968strip},  but the results are similar or worse than those obtained with MLM-PG. It is also known that even more sophisticated transport equation based models such as SA, SST, $k-\epsilon$ etc. fail to predict boundary layer parameters sufficiently accurate in the relaminarizing region \cite{spalart2000strategies}. 

For the comparative study between fully turbulent and quasi-laminar calculations in the following section, the results with MLM-PG turbulence model are compared with QLT for the relaminarizing region.

\subsection{Quasi-laminar layer}
In the QLT code, the equations for the inner and outer layers are solved separately as in NS73.
\subsubsection{Inner layer}
The equation for the inner layer Eqn. \ref{eq:inner} is solved directly using implicit finite difference method as against the Falkner-Skan or Thwaites method used in NS73. The initial profile for this laminar boundary layer is assumed to be the Pohlhausen cubic velocity profile with the pressure gradient and initial skin-friction  taken from Blasius skin-friction law. To solve the equation, the numerical schemes given in \cite{schetz1993boundary} are used with 251 points along the boundary layer. The wall shear stress ($\tau_w$) obtained from this inner layer solution is normalized with $0.5\rho U_e^2$ to obtain the skin-friction for the quasi-laminar region.     

\subsubsection{Outer layer}
To solve the outer layer equations, we follow the simple integral method suggested by NS73, where momentum and energy integral equations are used to obtain a first-order differential equation in the exponent of the power law velocity profile. This is acceptable as the overall solution of QLT is not very sensitive to the origin of the outer layer. In the present case, the initial power law velocity profile for the outer layer is obtained using a fit of the turbulent boundary layer profile (as calculated from the TBL code) at the corresponding location. The differential equation is then solved using the ODE45 solver available in Matlab 8.3. 

\begin{figure}[t]
\centering
\includegraphics[trim=0 0 0 0, clip, width=0.92\linewidth, angle=0]{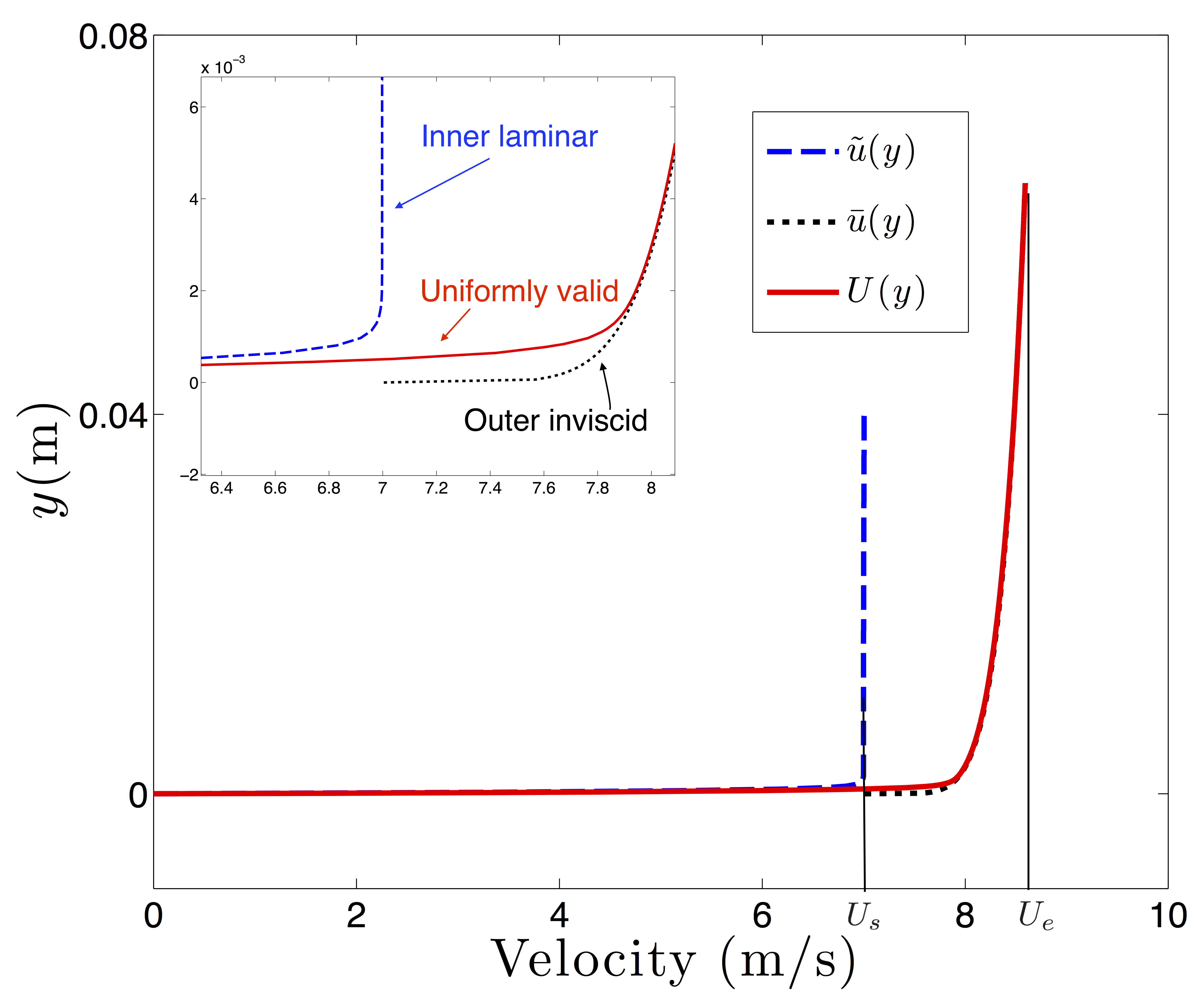}
\caption{Uniformly valid solution using 2-layer model}
\label{fig:uniform}
\end{figure}

\begin{figure}[t]
\centering
\includegraphics[trim=0 0 0 0, clip, width=0.9\linewidth, angle=0]{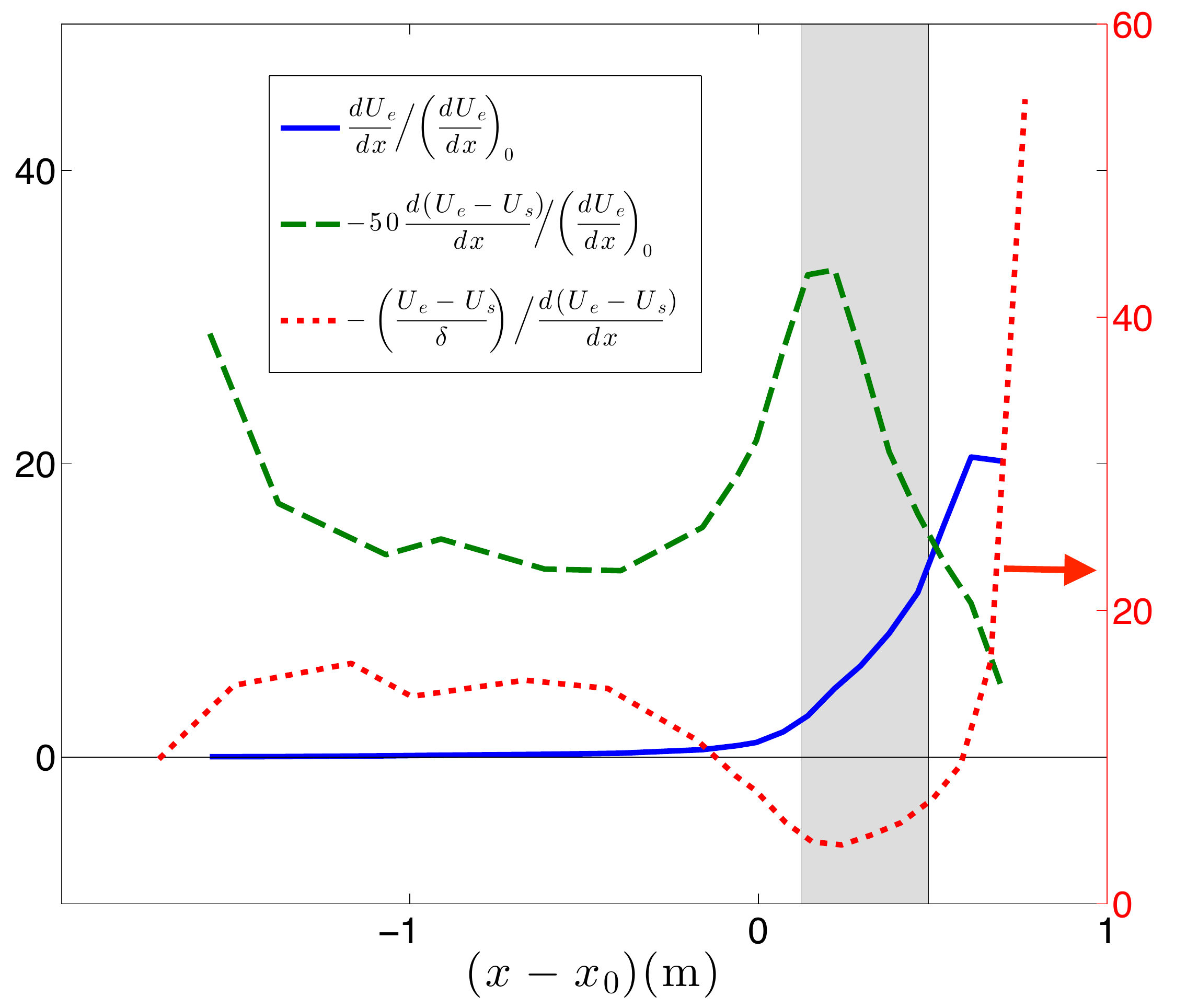}
\caption{Velocity gradients for BT flow. $(\textrm{d}U_e / \textrm{d}x)_0$ indicate streamwise velocity gradient at $x=x_0$}
\label{fig:gradients}
\end{figure}

Once the profiles for both the layers are known, the relevant boundary layer parameters of interest such as $\delta, \delta^{\star}, \theta, H$ etc. can be  obtained from the uniformly valid profile. Figure \ref{fig:uniform} shows the solutions for inner and outer layers along with the uniformly valid solution at a location $x-x_0 = 0.2$ in the relaminarizing region of BT-flow. 
\section{Results and discussion}\label{sec:results}
\begin{figure*}[t]
\centering
\includegraphics[width=1\linewidth]{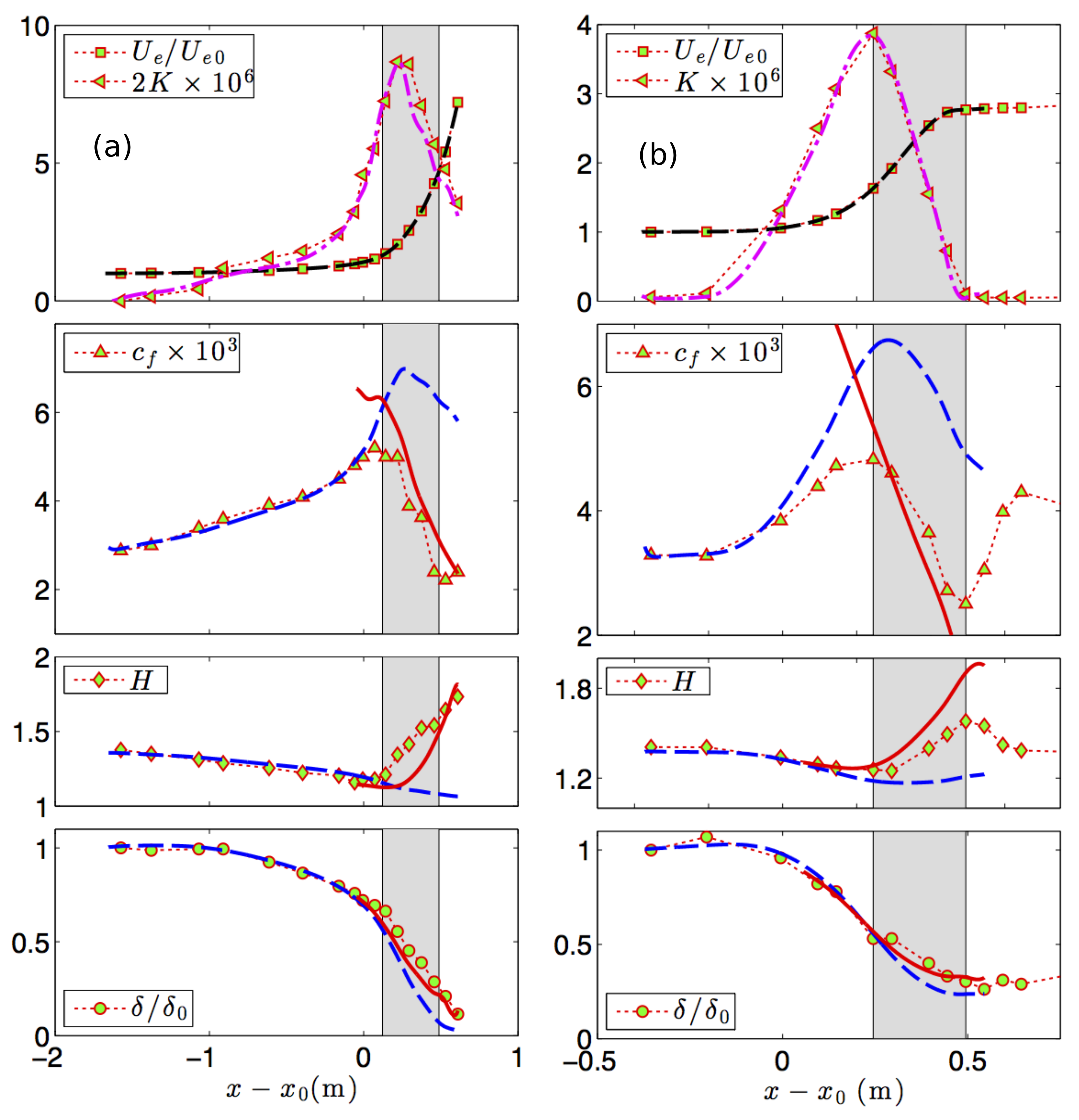}
\caption{Solutions for (a) BT (b) WF4 flows using TBL and QLT. Dashed and solid thick lines represent predictions with TBL and QLT respectively. Markers are experimental values.}
\label{fig:master}
\end{figure*}

In this section, we compare the results for BT-flow as obtained using QLT and TBL. Before proceeding further, it is important to verify that the  boundary layer (and hence quasi-laminar) assumptions  are valid in the high FPG region as the streamwise gradients are very high. Figure \ref{fig:gradients} compares the streamwise gradient with boundary layer shear for the full flow as well as outer flow. The substantial logitudinal strain rate in the flow is obvious from the fact that the non-dimensional velocity gradient $\textrm{d}U_e / \textrm{d} x$ goes as high as 20, whereas the maximum non-dimensional strain rate in the outer flow, which is $\textrm{d}(U_e-U_s) / \textrm{d} x$, is only 0.65. The shear $\partial U / \partial y$ in the outer layer (of  order  $(U_e - U_s) / \delta$), however, continues to be higher than this streamwise derivative  $\textrm{d} (U_e - U_s) / \textrm{d} x$ in the laminarized zone, by a factor between 4 and 20. This value can be considered sufficiently high to get approximate answers from boundary layer theory.

Figure \ref{fig:master}(a) shows the solution obtained for QLT as well as TBL for this flow. The top panel shows the distribution of external velocity as well as  pressure gradient parameter $K$ for both experiment and computation.  A cubic smoothing spline was used to fit $U_e$ obtained from experiment and a slight deviation in $K$ is expected. The deviation is unavoidable due to the necessity of using a smooth and differentiable velocity field. It should be mentioned that experimental data used here are hand-extracted from the figures given in the publication as we did not have access to digitized data. 

In the second panel from the top, the experimental values of $c_f$ are compared with the predictions of TBL and QLT.   For the region ($x<x_0$), where the flow is fully turbulent and pressure-gradient is very mild, the predictions with the TBL are very good; however in the strong pressure gradient region ($x>x_0$), $c_f$ is very poorly predicted. On the other hand, the predictions with the QLT are unacceptable in the fully turbulent region but they are dramatically better in the quasi-laminar region compared to TBL. The same trend is observed for shape factor as well as boundary layer thickness (bottom two panels), where predictions with QLT are much closer to experiment than TBL for the quasi-laminar region. It is expected that predictions in $c_f$ by QLT can be further improved by using a better initial profile. This work is currently in progress.

There is a small region in the boundary layer ($-0.05 < x -x_0< 1.05$) where none of these predictions, either by TBL or QLT, seem to be valid. This region was identified as transitional in \cite{narasimha1979relaminarization} and has been called  `island of ignorance' in \cite{sreenivasan1982laminarescent}. A careful interpolation based on  predictions from TBL and QLT using intermittency may make the predictions better in this region. This work is currently in progress. 

The superiority of QLT over TBL in the quasi-laminar region is supported by the other experiments mentioned in Table \ref{tab:study_relam}. For the sake of completeness, the predictions for another high $Re$ flow WF4 are presented in Fig. \ref{fig:master}(b). The observations made  regarding TBL and QLT above for the BT-flow can be repeated for this flow as well.    

\section{Conclusions}
A simple but robust quasi-laminar theory (QLT) has been proposed by \cite{narasimha1973relaminarization} to explain the later stages of relaminarization. QLT is based on a two-layer model:  a sheared outer stress-free inviscid layer under the effect of acceleration, and an inner viscous sub-boundary layer that is stabilized by the FPG and develops subsequently to satisfy the no-slip boundary condition. Earlier tests of this theory were mostly against low $Re$ experiments \cite{narasimha1973relaminarization, narasimha1979relaminarization}.  

In this paper, we assess the two-layer model based on QLT for recent experiments which have a relatively long turbulent boundary layer history (high initial $Re_{\theta0}$) and have been conducted with improved instrumentation and better control. For reasons stated in section \ref{sec:literature}, \cite{bourassa2009experimental} is considered for a detailed assessment of QLT. The basic principles behind this theory such as nearly frozen Reynolds stress and constant mass flux  in the outer layer are first assessed.

In the later stages of relaminarization, QLT provides a superior match with experimental results than turbulent boundary layer codes. The main limitation of the QLT is the choice of an effective virtual origin for the inner layer, which is due to lack of a precise definition of onset of relaminarization.

A new single integrated model, combining a turbulent boundary layer code with QLT but each in their respective regions of validity, has been proposed and is currently being evaluated.

\bibliographystyle{asmems4}
\bibliography{refer}

\begin{thebibliography}{10}

\bibitem{narasimha1973relaminarization}
Narasimha, R., and Sreenivasan, K.~R., 1973.
\newblock ``Relaminarization in highly accelerated turbulent boundary layers''.
\newblock {\em Journal of Fluid Mechanics, {\bf 61}}(03), pp.~417--447.

\bibitem{wilson1954convective}
Wilson, D., and Pope, J., 1954.
\newblock ``Convective heat transfer to gas turbine blade surfaces''.
\newblock {\em Proceedings of the Institution of Mechanical Engineers, {\bf
  168}}(1), pp.~861--876.

\bibitem{lagraff2007minnowbrook}
Halstead, D., Okiishi, T., and Wisler, D., 2007.
\newblock ``Boundary layer development on a turbine blade in a linear
  cascade''.
\newblock In {\em {Minnowbrook I: 1993 Workshop on End-Stage Boundary Layer
  Transition}}, J.~E. LaGraff, ed.

\bibitem{mukund2012multiple}
Mukund, R., Narasimha, R., Viswanath, P., and Crouch, J., 2012.
\newblock ``Multiple laminar-turbulent transition cycles around a swept leading
  edge''.
\newblock {\em Experiments in Fluids, {\bf 53}}(6), pp.~1915--1927.

\bibitem{massachusetts1956boundary}
MIT, G. T.~L., and Senoo, Y., 1956.
\newblock {\em Boundary Layer on the End Wall of a Turbine Nozzle Cascade}.

\bibitem{sternberg1954transition}
Sternberg, J., 1954.
\newblock The transition from a turbulent to a laminar boundary layer.
\newblock Tech. rep., DTIC Document.

\bibitem{narasimha1979relaminarization}
Narasimha, R., and Sreenivasan, K.~R., 1979.
\newblock ``Relaminarization of fluid flows''.
\newblock {\em Advances in Applied Mechanics., {\bf 19}}, pp.~221--309.

\bibitem{narasimha1983relaminarization}
Narasimha, R., 1983.
\newblock ``Relaminarization-magnetohydrodynamic and otherwise''.
\newblock {\em Progress in Astronautics and Aeronautics}(84), pp.~30--52.

\bibitem{bourassa2009experimental}
Bourassa, C., and Thomas, F., 2009.
\newblock ``An experimental investigation of a highly accelerated turbulent
  boundary layer''.
\newblock {\em Journal of Fluid Mechanics, {\bf 634}}, pp.~359--404.

\bibitem{warnack1998effects}
Warnack, D., and Fernholz, H., 1998.
\newblock ``{The effects of a favourable pressure gradient and of the Reynolds
  number on an incompressible axisymmetric turbulent boundary layer. Part 2.
  The boundary layer with relaminarization}''.
\newblock {\em Journal of Fluid Mechanics, {\bf 359}}, pp.~357--381.

\bibitem{sreenivasan1982laminarescent}
Sreenivasan, K.~R., 1982.
\newblock ``Laminarescent, relaminarizing and retransitional flows''.
\newblock {\em Acta Mechanica, {\bf 44}}(1-2), pp.~1--48.

\bibitem{launder1964laminarization}
Launder, B.~E., 1964.
\newblock Laminarization of the turbulent boundary layer by acceleration.
\newblock Tech. Rep.~77, MIT Gas Turbine Lab.

\bibitem{patel1965calibration}
Patel, V., 1965.
\newblock ``Calibration of the preston tube and limitations on its use in
  pressure gradients''.
\newblock {\em Journal of Fluid Mechanics, {\bf 23}}(01), pp.~185--208.

\bibitem{kline1967structure}
Kline, S., Reynolds, W., Schraub, F., and Runstadler, P., 1967.
\newblock ``The structure of turbulent boundary layers''.
\newblock {\em Journal of Fluid Mechanics, {\bf 30}}(04), pp.~741--773.

\bibitem{patel1968reversion}
Patel, V., and Head, M., 1968.
\newblock ``Reversion of turbulent to laminar flow''.
\newblock {\em Journal of Fluid Mechanics, {\bf 34}}(02), pp.~371--392.

\bibitem{narayanan1969criteria}
Narayanan, M.~B., and Ramjee, V., 1969.
\newblock ``On the criteria for reverse transition in a two-dimensional
  boundary layer flow''.
\newblock {\em Journal of Fluid Mechanics, {\bf 35}}(02), pp.~225--241.

\bibitem{blackwelder1972large}
Blackwelder, R.~F., and Kovasznay, L.~S., 1972.
\newblock ``Large-scale motion of a turbulent boundary layer during
  relaminarization''.
\newblock {\em Journal of Fluid Mechanics, {\bf 53}}(01), pp.~61--83.

\bibitem{escudier1998laminarisation}
Escudier, M., Abdel-Hameed, A., Johnson, M., and Sutcliffe, C., 1998.
\newblock ``Laminarisation and re-transition of a turbulent boundary layer
  subjected to favourable pressure gradient''.
\newblock {\em Experiments in Fluids, {\bf 25}}(5-6), pp.~491--502.

\bibitem{ichimiya1998properties}
Ichimiya, M., Nakamura, I., and Yamashita, S., 1998.
\newblock ``Properties of a relaminarizing turbulent boundary layer under a
  favorable pressure gradient''.
\newblock {\em Experimental Thermal and Fluid Science, {\bf 17}}(1),
  pp.~37--48.

\bibitem{piomelli2013numerical}
Piomelli, U., and Yuan, J., 2013.
\newblock ``Numerical simulations of spatially developing, accelerating
  boundary layers''.
\newblock {\em Physics of Fluids (1994-present), {\bf 25}}(10), p.~101304.

\bibitem{patwardhan2014effect}
Patwardhan, S.~S., 2014.
\newblock ``Effect of favourable pressure gradient on turbulence in boundary
  layers''.
\newblock PhD thesis, Indian Institute of Science, Bangalore, October.

\bibitem{krsnotes1974}
Sreenivasan, K.~R., 1974.
\newblock Notes on the experimental data on reverting boundary layers.
\newblock {Fluid Mech. Rep. 72 FM2, Dept. Aero. Engg.}, Indian Institute of
  Science, Bangalore.

\bibitem{schetz1993boundary}
Schetz, J.~A., 1993.
\newblock {\em Boundary layer analysis}.
\newblock Prentice Hall, NJ.

\bibitem{cebeci1968finite}
Cebeci, T., and Smith, A., 1968.
\newblock {\em A finite-difference solution of the incompressible turbulent
  boundary-layer equations by an eddy-viscosity concept}.
\newblock Douglas Aircraft Company.

\bibitem{clauser1956turbulent}
Clauser, F.~H., 1956.
\newblock ``The turbulent boundary layer''.
\newblock {\em Advances in Applied Mechanics, {\bf 4}}, pp.~1--51.

\bibitem{reichardt1951complete}
Reichardt, H., 1951.
\newblock ``Complete description of turbulent velocity profiles in smooth
  ducts''.
\newblock {\em ZAMM, {\bf 31}}, pp.~208--219.

\bibitem{spence2012development}
Spence, D., 2012.
\newblock ``The development of turbulent boundary layers''.
\newblock {\em Journal of the Aeronautical Sciences}.

\bibitem{head1960entrainment}
Head, M., 1960.
\newblock {\em Entrainment in the turbulent boundary layer}.
\newblock HM Stationery Office.

\bibitem{moses1968strip}
Moses, H.~L., 1968.
\newblock ``A strip-integral method for predicting the behavior of turbulent
  boundary layers''.
\newblock In Proc. AFOSR-IFP-Stanford Conference on Computation of Turbulent
  Boundary Layers.

\bibitem{spalart2000strategies}
Spalart, P.~R., 2000.
\newblock ``Strategies for turbulence modelling and simulations''.
\newblock {\em International Journal of Heat and Fluid Flow, {\bf 21}}(3),
  pp.~252--263.

\end{thebibliography}
\end{document}